\documentclass[prl,reprint,superscriptaddress]{revtex4-1}
\usepackage{graphicx}
\usepackage{amsmath}
\usepackage{amstext}
\usepackage{amssymb}
\usepackage{amsfonts}

\begin{document}
\title{Spectrum and Charge Ratio of Vertical Cosmic Ray Muons \\up to Momenta of 2.5 TeV/$c$}
\date{\today}

\author{M.~Schmelling}\email{Michael.Schmelling@mpi-hd.mpg.de}
\affiliation{Max-Planck-Institut f\"ur Kernphysik (MPIK), Heidelberg, Germany}
\author{N.O.~Hashim}
\affiliation{Department of Physics, Kenyatta University, Nairobi, Kenya}
\author{C.~Grupen}
\affiliation{University of Siegen, Faculty of Science and Technology,
Department of Physics, Siegen, Germany} 
\author{S.~Luitz} 
\affiliation{Stanford Linear Accelerator Center (SLAC), Stanford, California, USA}
\author{F.~Maciuc}
\affiliation{Max-Planck-Institut f\"ur Kernphysik (MPIK), Heidelberg, Germany}
\author{A.~Mailov} 
\affiliation{University of Siegen, Faculty of Science and Technology,
Department of Physics, Siegen, Germany} 
\author{A.-S.~M\"uller}
\affiliation{Karlsruher Institut f\"ur Technologie (KIT), Karlsruhe, Germany}
\author{H.-G.~Sander}
\affiliation{Institut f\"ur Physik, Universit\"at Mainz, Mainz, Germany}
\author{S.~Schmeling}
\affiliation{European Organization for Nuclear Research (CERN), Geneva, Switzerland}
\author{R.~Tcaciuc}
\affiliation{University of Siegen, Faculty of Science and Technology,
Department of Physics, Siegen, Germany} 
\author{H.~Wachsmuth}\altaffiliation{deceased}
\affiliation{European Organization for Nuclear Research (CERN), Geneva, Switzerland}
\author{K.~Zuber}
\affiliation{Institut f\"ur Kern- und Teilchenphysik, TU Dresden, Dresden, Germany}

\collaboration{The CosmoALEPH Collaboration}
\noaffiliation

\begin{abstract}
  \centerline{\em Dedicated to the memory of Horst Wachsmuth}
\vspace*{2mm}
  The ALEPH detector at LEP has been used to measure the momentum
  spectrum and charge ratio of vertical cosmic ray muons underground.
  The sea-level cosmic ray muon spectrum for momenta up to $2.5$\,TeV/$c$\/
  has been obtained by correcting for the overburden
  of 320 meter water equivalent (mwe). The results are compared with 
  Monte Carlo models for air shower development in the atmosphere.
  From the analysis of the spectrum the total flux and the spectral index of
  the cosmic ray primaries is inferred. The charge ratio suggests a dominantly 
  light composition of cosmic ray primaries with energies up to $10^{15}$\,eV.
\end{abstract}

\maketitle

The sea level cosmic ray muon spectrum resulting from the interaction
of primary cosmic rays with the atmosphere probes multi-particle
production in hadronic interactions up to and beyond LHC energies.
In addition, the muon charge ratio is sensitive to the fraction of 
heavy nuclei in the primary cosmic ray beam.

In the past many measurements were based on simple detection
devices at the earth's surface, see e.g. \cite{Hebbeker} and
references therein. Additional information can be
obtained from the highly sophisticated detectors used in 
particle physics \cite{L3,CMS,Opera,Minos}. The fact that those 
detectors are placed under ground offers the 
additional advantage of a natural muon filter and a momentum 
cutoff which emphasizes the high energy part of the spectrum.
Here we present a measurement of the vertical muon spectrum 
and charge ratio performed with the ALEPH detector at LEP,
which was located at a depth of 320 mwe.

The ALEPH detector is described in detail in \cite{Aleph}. The
study presented here uses mainly the large time projection chamber
(TPC), the main tracking device of the ALEPH detector. The data were
taken in 1999 during one week of dedicated running with a cosmic
trigger using information from the hadron calorimeter (HCAL).
Background from interactions in the iron of the HCAL and
electromagnetic calorimeter (ECAL) can be reliably identified, have 
been shown to be well described by the detector simulation \cite{Maciuc} 
and are taken into account in the Monte Carlo corrections.

\begin{figure}[htb]
\centering
  \includegraphics[width=0.44\textwidth]{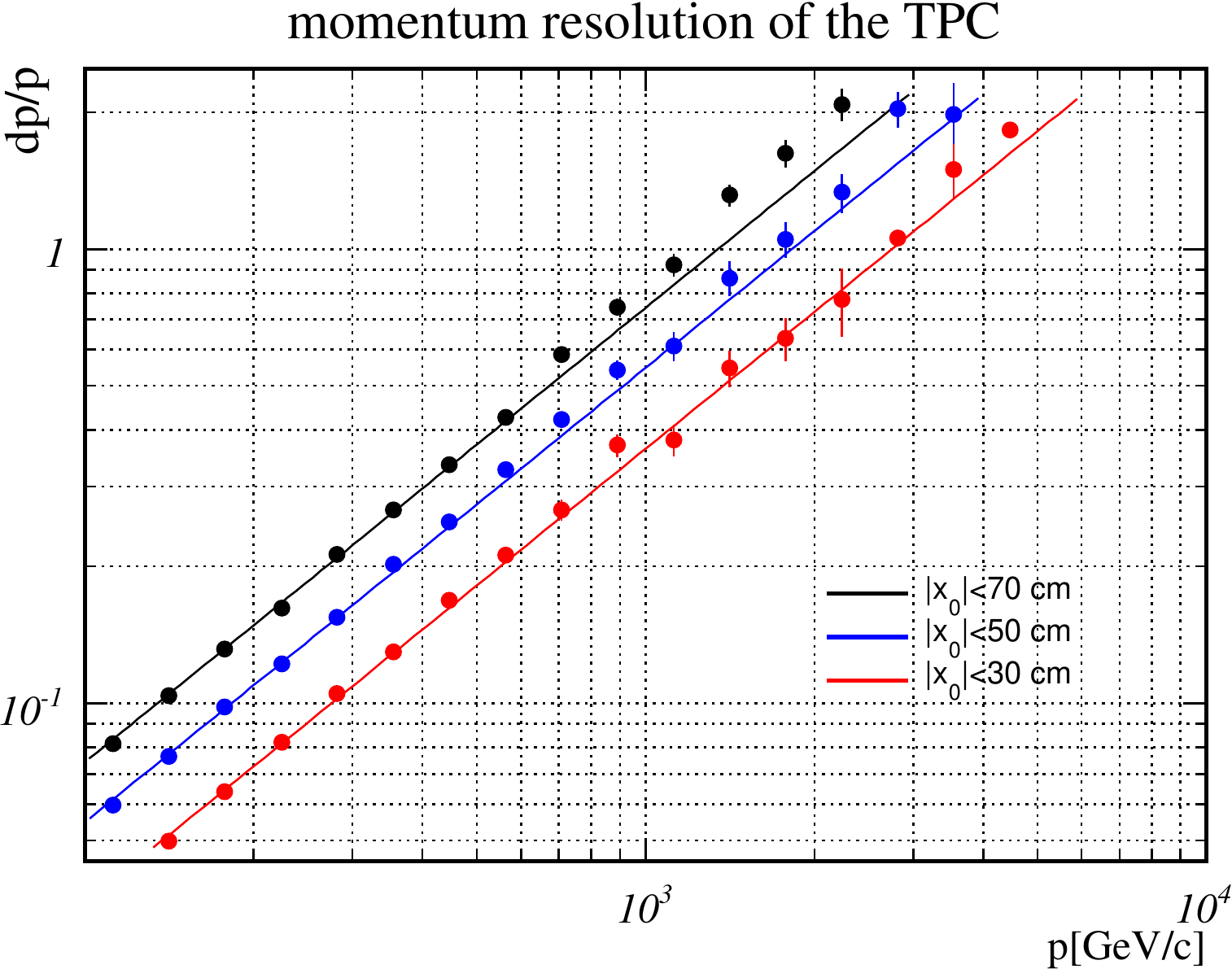}
 \caption{Momentum resolution as a function of the momentum
          for three cuts on the fiducial area of the TPC.}
\label{fig:pmax}
\end{figure}

The coordinate system used in the analysis is the standard ALEPH
coordinate system. It is a right-handed system with the $x$-direction
pointing towards the center of LEP, the $y$-direction being vertically 
upwards and $z$\/ along the beam-direction. The center of the detector
is at $x=y=z=0$. Muon tracks are selected by the requirement of having
at least $27$\/ TPC hits out of a maximum of $42$, a reconstructed
momentum $p>10$\,GeV/$c$\/ and a polar angle with respect to the vertical 
$\theta<20^\circ$. The cut on the number of TPC hits guarantees precise momentum 
reconstruction, the minimum momentum cut ensures that the muon has enough
energy to pass through the entire barrel of the calorimeter and
thus full trigger efficiency. Both cuts reject low energy background 
from secondary interactions. The quality of the reconstruction is 
demonstrated by the fact that such details of the geometrical 
structure as the access shaft of the ALEPH pit and the flux
shadowing by the Jura mountains are visible in the arrival 
directions of the incident muons. For the final analysis additional 
fiducial cuts were applied. Azimuthal angles (measured around the vertical) 
of $\pm 10^\circ$\/ with respect to the direction of the underground access 
shaft were cut out to mask the enhanced muon flux from this direction.
In addition, the muon tracks were required to pass through
the central plane $y=0$\/ in the range $|x_0|<50$\,cm, $|z_0|>50$\,cm
and $|z_0|<170$\,cm. These cuts make sure that the selected tracks 
are fully contained in one half of the TPC and pass through the 
full diameter of the chamber, optimizing momentum resolution
and minimizing systematic effects due to boundary effects and
the inversion of the drift field at the HV-membrane at $z=0$.
The momentum resolution $dp/p$\/ of the TPC is shown in 
Fig.~\ref{fig:pmax} as a function of $p$\/ for three values
of the fiducial cut on $x_0$. One clearly sees how the 
performance improves towards the central region. The nominal
cut value $|x_0|<50$\,cm is chosen as a compromise between good 
resolution and still high statistics. Here the maximum measurable 
momentum is around $2$\,TeV/$c$. Extrapolated to the surface, the 
accessible momentum range is up to 2.5\,TeV/$c$.

The raw distributions for spectrum and charge ratio were corrected 
using momentum dependent correction factors obtained from a full simulation 
of single muons passing through the ALEPH detector from above. For this 
simulation, at a depth of 320 mwe, an energy spectrum proportional to $E^{\gamma}$\/ 
was assumed, with $\gamma=-3$. The correction factors, parameterized by
4th order polynomials as a function of $\lg p$, include effects due
to electromagnetic interactions of muons in the detector material,
multiple scattering in the calorimeters and finite momentum resolution
of the TPC. For momenta from $10$\/ to $40$\,GeV/$c$\/ the correction 
factors for the spectrum vary from  values around $2$\/ to $1.2$. Towards
higher momenta they decrease slowly, falling below unity around $1$\,TeV/$c$. 
For the charge ratio no trend is observed. With average corrections
below $1$\,\%, the analysis of the charge ratio was performed
without explicit detector corrections. In the end a global $1$\,\%
systematic error, correlated between all bins, was assigned. The 
nominal results for the muon momentum spectrum and charge ratio at 
the surface were then determined by propagating the measured momenta
to the surface, using a parameterization of the form $dE/dx=a+bE$\/ 
with energy dependent coefficients $a$\/ and $b$ \cite{Nadir}.
The continuous energy loss is approximated by performing the propagation
in five discrete steps, which has been checked to be within $1\%$ 
of the continuous treatment.
An overburden of $306$\,mwe for $\phi<180$\,deg (towards lake Geneva) and 
$326$\,mwe  for $\phi>180$\,deg (towards Jura mountains) was assumed, which 
takes into account the surface topography close to the ALEPH experiment. 
After propagation  to the surface the spectrum measured up to zenith 
angles of 20\,degrees was corrected to the purely vertical spectrum 
by a momentum dependent factor $I(p,0)/I(p,\theta)$\/ using the
parameterization of the muon flux in reference \cite{Bogdanova}.
The correction is in the range $1$--$3\%$. The absolute flux finally is 
obtained from
\begin{equation}
   \Phi_{\mu} = \frac{N^{\tt corr}_{\mu}}
                     {T \cdot S \cdot \Omega \cdot \Delta p} \;.
\end{equation}
Here $N^{\tt corr}_\mu$\/ is the number of muons measured after all 
corrections in the momentum interval $[p,p+\Delta p]$, $S$\/ is the 
fiducial area in the TPC and $\Omega$\/ the solid angle accepted for 
the measurement. The uptime $T$\/ is estimated from an independent 
measurement \cite{ANA01} of the CosmoALEPH HCAL trigger rate 
and the number of muons per unit area observed during the data taking
as $T=(3.52\pm 0.32)\cdot 10^5$\,s where the error is dominated by 
systematics estimated from spatial inhomogeneities of the observed muon 
flux through the central plane of the TPC. Details of the analysis 
are described in references \cite{Nadir,ANA02}. The 
total normalization uncertainty is 9.1\%.

For the determination of the systematic uncertainties of the results
the analysis was repeated $10^4$\/ times, randomly varying global
settings of the analysis and fluctuating the individual measurements
within their uncertainties. With $D^{nom}_i$\/ the result from the 
nominal analysis, a covariance matrix describing the systematic 
uncertainties and correlations between all bins was constructed 
from those pseudo-experiments by 
\begin{equation}
    C_{ij} = \langle(D_i - D^{nom}_i)(D_j- D^{nom}_j)\rangle \;.
\end{equation}

Individual tracks were subjected to smearing of $1/p$\/ according to the 
estimated error from the track fit, plus direction and position smearing
due to multiple scattering in the overburden and the finite resolution 
of the TPC as parameterized from Monte Carlo simulations. 
Global variations between different pseudo-experiments
which have been considered are the spectral index in the calculation 
of the detector corrections, $\gamma\in\{-2.7,-3.0,-3.3\}$, the use of
an additional quality cut $dp/p<1$\/ in the track selection, omitting 
the zenith angle correction, or, in order to probe the sensitivity to 
the simple energy loss model, doing the propagation through the 
overburden in a single step. Continuous global parameters entering 
the analysis have been randomly varied between pseudo-experiments by
assuming a uniform distribution within their assumed uncertainties. 
The overburden was varied by $\pm 3.4$\,mwe on the lake side
and $\pm13.4$\,mwe on the Jura side, where the difference accounts 
for uncertainties in the exact effects of the surface topology.
Momentum dependent biases on position, direction and momentum measurements
in the TPC are assumed to be taken into account by the Monte Carlo simulation. 
Nevertheless, in order to probe the sensitivity to biases seen in the 
simulation, these were parameterized and added with a scale factor randomly 
chosen from a uniform distribution in the range $[-1,1]$, i.e.\ covering the 
range between zero and twice the estimated bias. In the same way also the 
impact of the uncertainty of the global momentum scale 
$\sigma(1/p)=3.7\cdot 10^{-5}/$GeV/$c$\/ \cite{Wiedenmann} was studied.

\begingroup
\begin{table}[t]
\centering
\footnotesize
\begin{tabular}{|c|l|c|}
\hline
 $\bar{p}$ [GeV/$c$]
& spectrum [(GeV/$c$ s cm$^2$ sr)$^{-1}]$
& charge ratio \\
\hline
 112 & $(1.959 \pm 0.018 \pm 0.227)\cdot 10^{-7}$  & $1.252 \pm 0.024 \pm 0.046$\\
 141 & $(9.004 \pm 0.108 \pm 0.648)\cdot 10^{-8}$  & $1.293 \pm 0.030 \pm 0.025$\\
 178 & $(4.519 \pm 0.063 \pm 0.208)\cdot 10^{-8}$  & $1.259 \pm 0.036 \pm 0.017$\\
 224 & $(2.202 \pm 0.040 \pm 0.062)\cdot 10^{-8}$  & $1.271 \pm 0.046 \pm 0.028$\\
 282 & $(1.072 \pm 0.025 \pm 0.042)\cdot 10^{-8}$  & $1.239 \pm 0.057 \pm 0.033$\\
 355 & $(5.068 \pm 0.152 \pm 0.157)\cdot 10^{-9}$  & $1.348 \pm 0.080 \pm 0.048$\\
 447 & $(2.373 \pm 0.090 \pm 0.071)\cdot 10^{-9}$  & $1.541 \pm 0.121 \pm 0.085$\\
 562 & $(1.204 \pm 0.058 \pm 0.081)\cdot 10^{-9}$  & $1.373 \pm 0.133 \pm 0.101$\\
 708 & $(5.745 \pm 0.034 \pm 0.392)\cdot 10^{-10}$ & $1.243 \pm 0.153 \pm 0.127$\\
 891 & $(2.745 \pm 0.214 \pm 0.209)\cdot 10^{-10}$ & $1.547 \pm 0.248 \pm 0.168$\\
1122 & $(1.209 \pm 0.126 \pm 0.102)\cdot 10^{-10}$ & $1.785 \pm 0.388 \pm 0.369$\\
1413 & $(6.741 \pm 0.829 \pm 0.991)\cdot 10^{-11}$ & $1.361 \pm 0.339 \pm 0.368$\\
1778 & $(2.217 \pm 0.419 \pm 0.871)\cdot 10^{-11}$ & $0.648 \pm 0.251 \pm 0.805$\\
2239 & $(1.536 \pm 0.307 \pm 0.548)\cdot 10^{-11}$ & $1.495 \pm 0.610 \pm 0.698$\\
\hline
\end{tabular}
\caption{Vertical muon spectrum and charge ratio at sea level
for momenta between 100\,GeV/$c$ and 2.5\,TeV/$c$. The first 
column gives the momentum  at the bin center,
the second and third columns are the vertical flux and the charge
ratio, respectively. The first error is statistical, the 
second one the systematic uncertainty of the results. In addition 
there is a global normalization uncertainty of 9.1\% for the 
spectrum. The correlation matrix of the systematic errors is 
given in a supporting document to this analysis \cite{ANA02}.}
\label{tab:data}
\end{table}
\endgroup

\begin{figure}[t]
\centering
\includegraphics[width=0.44\textwidth]{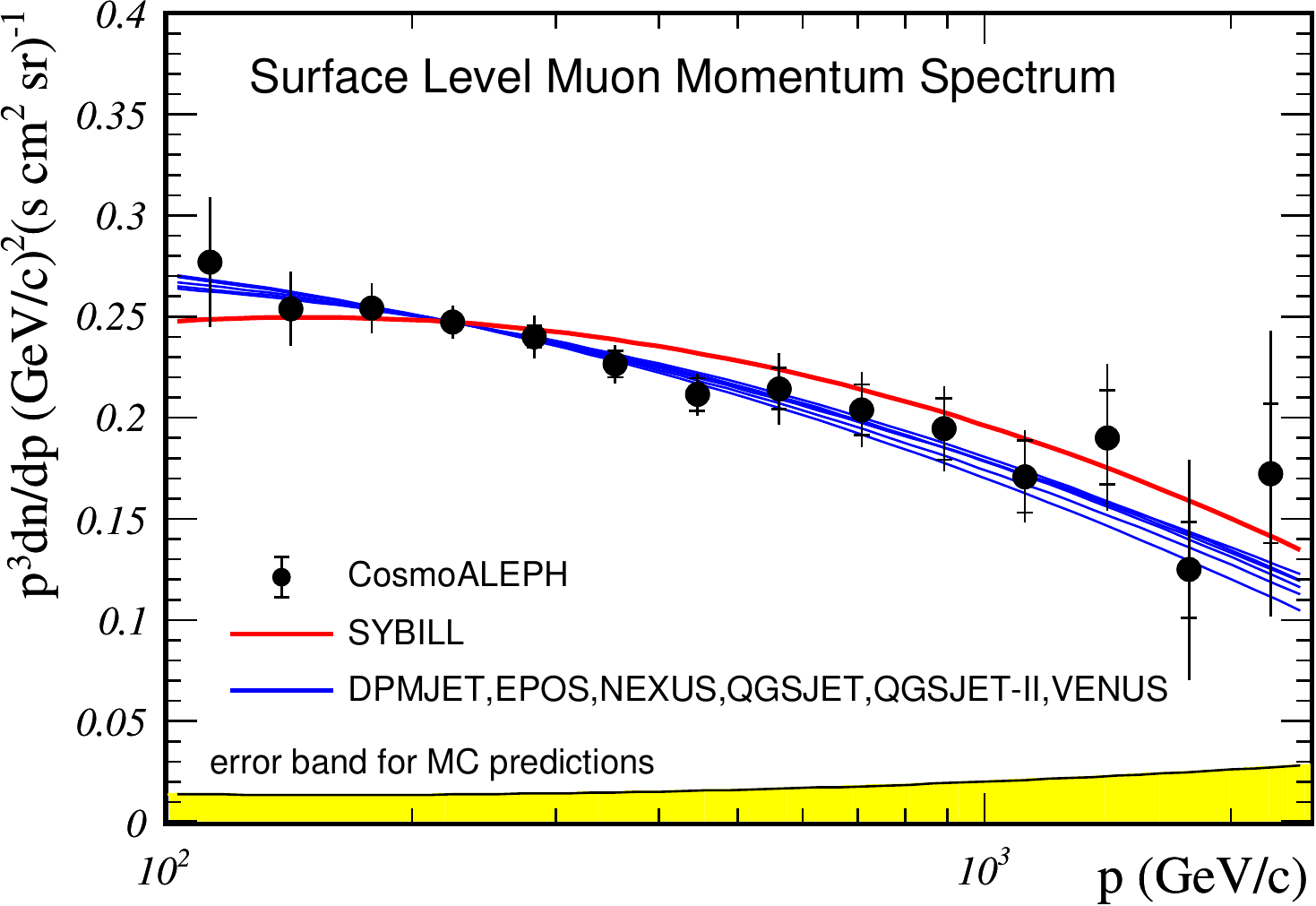}
\includegraphics[width=0.44\textwidth]{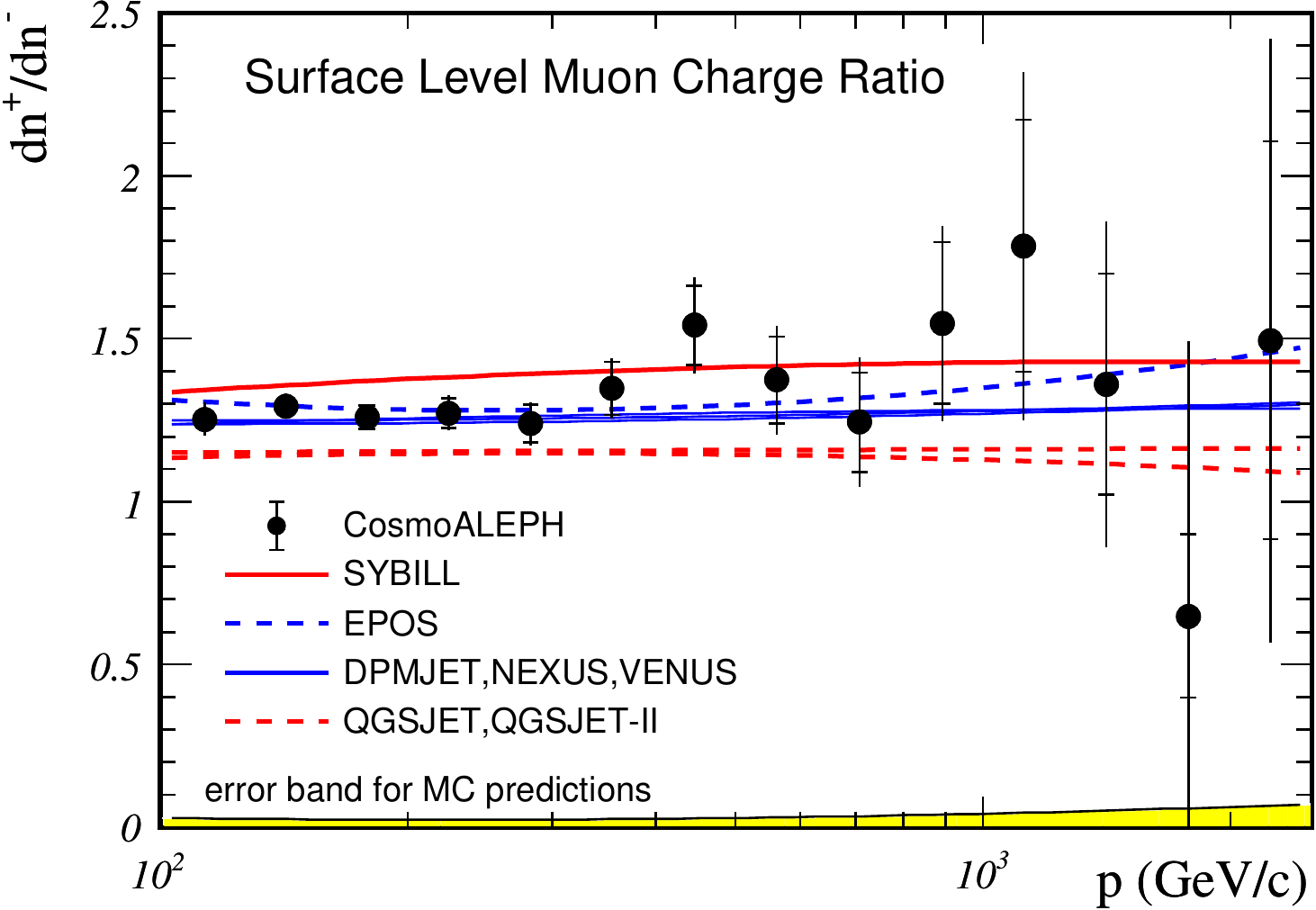}
\caption{Measured momentum spectrum and charge ratio compared 
         to best fit Monte Carlo predictions. At the bottom the 
         error bands of the Monte-Carlo curves are indicated.}
\label{fig:mccomp}
\end{figure}

The CosmoALEPH results for the vertical muon spectrum and the charge
ratio are listed in table \ref{tab:data} and displayed together with
predictions from different interactions models in Fig.~\ref{fig:mccomp}.
The spectrum agrees with the global parameterization given
in reference \cite{Hebbeker}. Taking the full covariance matrix of the 
result into account, one finds a $\chi^2$-confidence level $p=18.6\,\%$.
At large momenta the CosmoALEPH measurements are also in good agreement 
with the results by the MARS collaboration \cite{MARS}, which was 
historically the first to measure momenta beyond $1$\,TeV/$c$.

Within errors the measurements of the charge ratio are consistent with 
being independent of the momentum, but also with an increase at large
energies as observed e.g. in \cite{CMS,Opera,Minos}. Fitting a constant 
value gives a $\chi^2$-confidence level $p=0.85$\/ and the charge ratio
$R_{\mu}=1.289\pm 0.022$, which is in good agreement with the result from 
\cite{Hebbeker} based on momenta up to $316$\,GeV/$c$. The CosmoALEPH
measurement reaches up to $2.5$\,TeV/$c$, albeit with large uncertainties 
for momenta above $1$\,TeV/$c$.

The CosmoALEPH measurements were compared with Monte Carlo simulations
using the models
{\tt DPMJET}\cite{DPMJET}, 
{\tt VENUS}\cite{VENUS}, 
{\tt NEXUS}\cite{NEXUS}, 
{\tt EPOS}\cite{EPOS},
{\tt QGSJET}\cite{QGSJET}, 
{\tt QGSJETII}\cite{QGSJETII} and 
{\tt SYBILL\cite{SYBILL}}
from the CORSIKA program suite, version 6.375 
\cite{CORSIKA}. Studies of different parameterizations of the 
primary spectrum \cite{Forti} showed that the data are reasonably 
well described by the Constant Mass Composition model (CMC),
both in shape and absolute normalization. The momentum range 
covered by the CosmoALEPH data is found to be insensitive to 
the existence of the knee in the primary spectrum. The observed
charge ratio depends mainly on the average charge per nucleon 
of the primary particle initiating a shower, i.e.\ it is sensitive
to the fraction of heavy particles in the primary spectrum, but 
does not provide information about the chemical composition of 
those nuclei. The momentum spectrum of cosmic ray muons is sensitive 
to the total flux and the spectral index of the primary spectrum.

To determine the fraction of heavy nuclei $f_h$, total flux and  
spectral index $\gamma$, Monte Carlo predictions using simple power 
laws 
\begin{equation}
\label{model}
    \frac{dn}{dE_h} = K\, f_h\, E^{-\gamma}
    \quad\mbox{and}\quad
    \frac{dn}{dE_p} = K\, (1-f_h)\, E^{-\gamma}
\end{equation}
for the heavy particles and protons were fitted to the data. 
For heavy nuclei the cocktail of the CMC-model with about $30\%$\,He-, 
$20\%$\,N-, $30\%$\,Mg- and  $20\%$\,Fe-nuclei was assumed.
Technically, in the fit the existing simulations were reweighted 
to conform to eq.\,\ref{model}. The parameters $f_h$\/ and $\gamma$\/ were 
scanned in the range $2.3<\gamma<2.9$\/ and $0<f_h<1$. At each point 
the normalization $K$\/ of the spectrum was adjusted such that the model 
reproduces the best measured point of the momentum spectrum. Then 
the total $\chi^2$\/ for the remaining 27 degrees of freedom from 
spectrum and charge ratio was calculated. The two distributions were 
assumed to be independent, but the full correlation matrix within 
each was properly taken into account. The best fit results are 
shown in Fig.~\ref{fig:mccomp}.

\begin{figure}[t]
\centering
\includegraphics[width=0.44\textwidth]{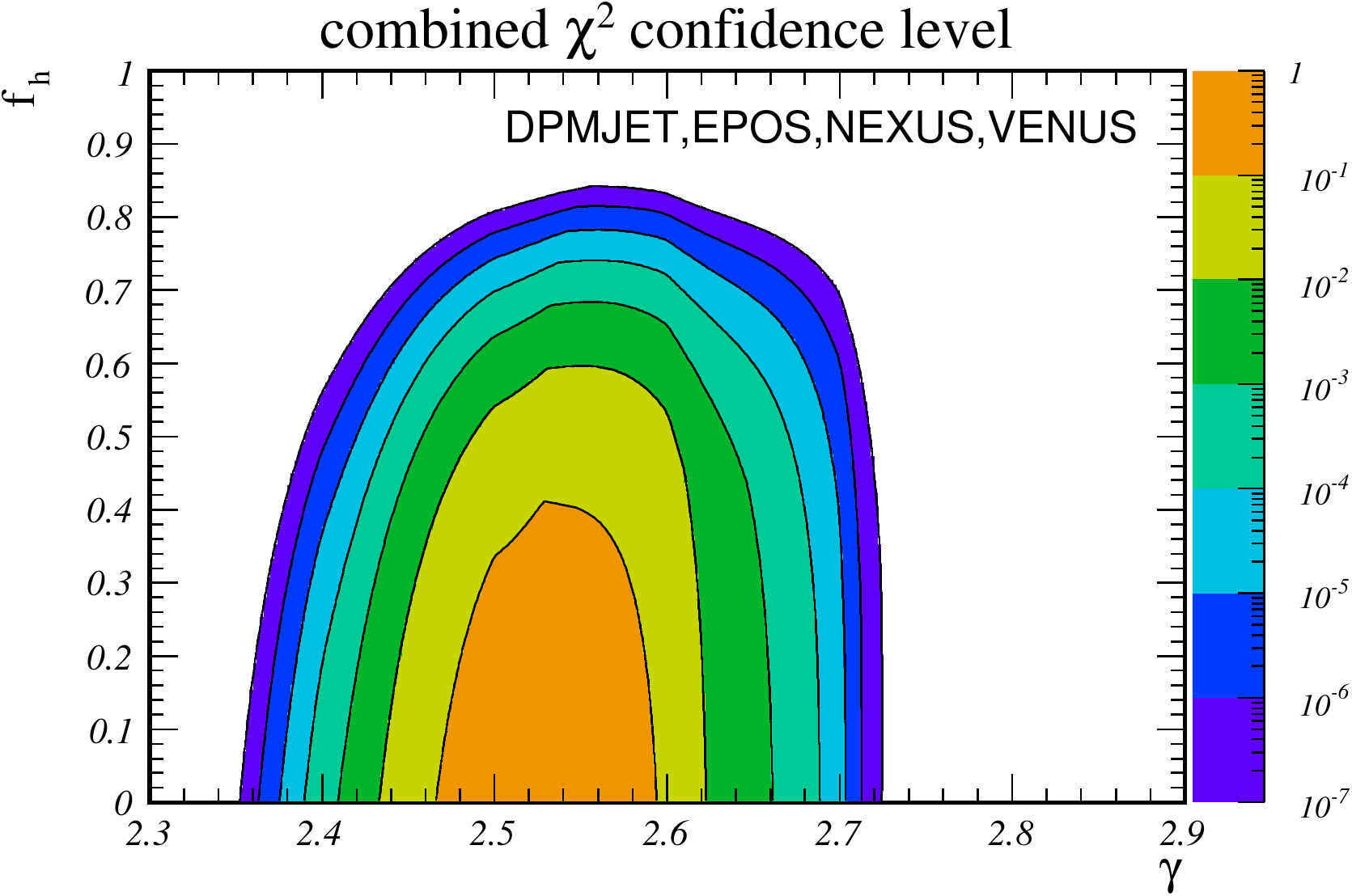}
\caption{$\chi^2$-confidence level contours as a function of $\gamma$\/ 
         and $f_h$. The plot shows the combined confidence level 
         contours from the models with a good global fit.}
 \label{fig:chi2cl}
\end{figure}

One finds that all models except {\tt SYBILL} achieve a good 
description of the shape of the muon spectrum. Good global fits 
are obtained with the {\tt DPMJET, EPOS, NEXUS} and {\tt VENUS} 
models. The others fail to reproduce the observed charge ratio.
The four models with a good global fit favor a predominantly light 
composition of cosmic primaries. For a quantitative 
conclusion the results from those models were combined such that 
for every point in the $(\gamma,f_h)$\,plane the smallest of the 
four $\chi^2$-confidence levels was taken as the combined one. This 
procedure yields regions in the parameter plane, where all models 
under consideration are consistent with the data with at least 
the specified confidence level. The result is shown in Fig.~\ref{fig:chi2cl}.
Taking the region with common confidence level $p>0.1$\/ finally 
yields $2.47 <\gamma < 2.59$ and $f_h < 0.41$.

The flux normalization obtained from the fit is strongly correlated 
to $f_h$\/ and $\gamma$. Expressed through the total flux $\Phi_2$\/ 
of cosmic ray primaries with energies above $10^2$\,GeV and
considering only the region with common confidence level $p>0.1$, 
one finds $\Phi_2 = 3.68 \pm 0.51 \pm 0.33$\,m$^{-2}$s$^{-1}$sr$^{-1}$. 
The central value is the average of the results from the 
models with a good fit to the data, the first error the RMS 
spread of those results and the second one the $9.1\%$\/ 
normalization uncertainty of the CosmoALEPH measurement.
The statistical errors of the result are negligible.

In summary, we have presented a precision measurement of the spectrum 
and charge ratio of vertical cosmic ray muons at the surface in 
the momentum range from 100\,GeV/$c$ to 2.5\,TeV/$c$.
The charge ratio is sensitive to the fraction of heavy primaries. 
Based on simulations with different interaction models, the fraction 
of heavy nuclei is inferred to be $f_h<41\%$. The preferred
spectral index for the primary spectrum up to energies of $10^{15}$\,eV 
is found to be in the range from $2.47<\gamma<2.59$, and the flux of 
cosmic ray primaries with energy above $10^2$\,GeV is determined as
$\Phi_2 = 3.74 \pm 0.61$\,m$^{-2}$s$^{-1}$sr$^{-1}$.

The authors gratefully acknowledge the help of the ALEPH
collaboration, in particular Markus Frank and Beat Jost, 
as well as Alois Putzer, Bertram Rensch, and Thomas Ziegler 
in doing the measurements. The analysis of 
the CosmoALEPH experiment has been supported by the 
Deutsche Forschungsgemeinschaft under grant DFG/Gr/1796/1-3.

\end{document}